\documentclass[aps,prx,onecolumn,superscriptaddress,11pt]{revtex4-2} %linenumbers
\usepackage{graphicx}
\usepackage{units}
\usepackage{amssymb}
\usepackage{braket}
\usepackage{bm}
\usepackage[colorlinks=true, linkcolor=black, citecolor=black, urlcolor=black]{hyperref}

\usepackage{ulem}

\begin{document}
\title{A hole spin qubit in a fin field-effect transistor above 4 kelvin}

\author{Leon C. Camenzind}
\thanks{These authors contributed equally to this work;}

\author{Simon Geyer}
\thanks{These authors contributed equally to this work;} 
\affiliation{Department of Physics, University of Basel, Klingelbergstrasse 82, CH-4056 Basel, Switzerland}

\author{Andreas Fuhrer}
\affiliation{IBM Research-Z\"urich, S\"aumerstrasse 4, CH-8803 R\"uschlikon, Switzerland}

\author{Richard J. Warburton}

\author{Dominik M. Zumb\"uhl}
\email{e-mail: dominik.zumbuhl@unibas.ch, andreas.kuhlmann@unibas.ch}
\affiliation{Department of Physics, University of Basel, Klingelbergstrasse 82, CH-4056 Basel, Switzerland}

\author{Andreas V. Kuhlmann}
\email{e-mail: dominik.zumbuhl@unibas.ch, andreas.kuhlmann@unibas.ch}
\affiliation{Department of Physics, University of Basel, Klingelbergstrasse 82, CH-4056 Basel, Switzerland}
\affiliation{IBM Research-Z\"urich, S\"aumerstrasse 4, CH-8803 R\"uschlikon, Switzerland}

\date{\today}

\begin{abstract}
The greatest challenge in quantum computing is achieving scalability. Classical computing previously faced a scalability issue, solved with silicon chips hosting billions of fin field-effect transistors (FinFETs). These FinFET devices are small enough for quantum applications: at low temperatures, an electron or hole trapped under the gate serves as a spin qubit. Such an approach potentially allows the quantum hardware and its classical control electronics to be integrated on the same chip. However, this requires qubit operation at temperatures above \unit[1]{K}, where the cooling overcomes heat dissipation. Here, we show that silicon FinFETs can host spin qubits operating above \unit[4]{K}. We achieve fast electrical control of hole spins with driving frequencies up to \unit[150]{MHz}, single-qubit gate fidelities at the fault-tolerance threshold, and a Rabi oscillation quality factor greater than 87. Our devices feature both industry compatibility and quality, and are fabricated in a flexible and agile way that should accelerate further development.
\end{abstract}
\maketitle

Quantum dot (QD) spin qubits \cite{Loss1998} in silicon have potential applications in large-scale quantum computation \cite{Fowler2012}, due to their long coherence times \cite{Veldhorst2014} and high quality factors \cite{Takeda2016,Yoneda2017,Yang2019}, as well as the fact that complementary metal-oxide-semiconductor (CMOS) manufacturing processes \cite{Maurand2016,Kuhlmann2018,Geyer2021} can be used to create dense arrays of interconnected spin qubits \cite{Vandersypen2017,Veldhorst2017}. Inspired by conventional integrated circuits, on-chip integration of the classical control electronics with the qubit array has been suggested as a way to overcome the challenges in wiring up large numbers of multi-terminal QD devices \citep{Franke2019}. However, since the electronics produce heat, the amount of control functionality that can be implemented depends on the available cooling power. Therefore, it is beneficial to be able to operate qubits at temperatures greater than \unit[1]{K}, where cooling power is orders of magnitude higher than at mK temperatures \cite{Petit2020,Yang2020}. For example, Intel’s cryogenic control chip, which is known as Horse Ridge, works at \unit[3]{K} \cite{Xue2021}.

Spin qubits come in two distinct forms: electron \cite{Veldhorst2014,Kawakami2014,Takeda2016,Yoneda2017,Petit2020,Yang2020} and hole \cite{Maurand2016,Watzinger2018,Hendrickx2020,Hendrickx2021,Froning2021}. With electrons, an artificial spin-orbit interaction (SOI) can be engineered by equipping the qubit with a micromagnet \cite{Kawakami2014,Takeda2016,Yoneda2017}. Conversely, hole spins experience a strong intrinsic SOI \cite{Kloeffel2018}. All-electrical spin control is achieved via electric-dipole spin resonance (EDSR) \cite{Golovach2006,Nowack2007,NadjPerge2010,Voisin2015,Crippa2018}, where an applied oscillating electric field induces spin rotations. Compared to electrons, holes can reduce device complexity, which benefits scalability, because no additional device components are required to generate a SOI. Furthermore, with holes in silicon nanowires or FinFETs \cite{Auth2012,Auth2017,Lansbergen2008}, the SOI can be exceptionally strong and fully tunable, creating a switchable coupling strength and a way to mitigate the effects of charge noise \cite{Kloeffel2018,Bosco2021,Froning2021}. Moreover, hole spins are better protected against nuclear spin noise due to their weak hyperfine interaction \cite{Prechtel2016}. 

Recently, electron spin qubits operating up to \unit[1.5]{K} have been demonstrated \cite{Petit2020,Yang2020}. In this article, we report hole spin qubits working at 1.5 to \unit[5]{K}: that is, in a temperature range where the thermal energy is much larger than the qubit level splitting and cryogenic control electronics can be operated \cite{Xue2021}. The hole spin qubits are integrated in silicon FinFET devices that are created using standard CMOS fabrication techniques, including self-aligned gates and chemically-selective plasma etches instead of lift-off processes \cite{Kuhlmann2018,Geyer2021}. In addition, a high degree of process flexibility and a short turnaround are achieved by using electron-beam instead of advanced optical lithography \cite{Zwerver2021}. The fin provides a one-dimensional confinement for the holes, enabling fast and electrically tunable effective spin-1/2 qubits \cite{Kloeffel2018,Bosco2021,Froning2021}. We demonstrate EDSR-based spin control with Rabi frequencies up to $\unit[150]{MHz}$ and voltage-tunable qubit frequencies, a feature employed to implement z-rotations as fast as $\unit[45]{MHz}$. We also show spin rotations around the $x$- and $y$-axis of the Bloch sphere with a single-qubit gate fidelity of $\unit[98.9]{\%}$ at \unit[1.5]{K}. A high robustness against temperature allows for qubit operation above the boiling point of liquid $^4$He, albeit with a reduced dephasing time $T_2^*$ compared to $\unit[1.5]{K}$, which is consistent with an observed whitening of the spectral noise density on increasing temperature.

\begin{figure}
\centering
\includegraphics[width=\textwidth]{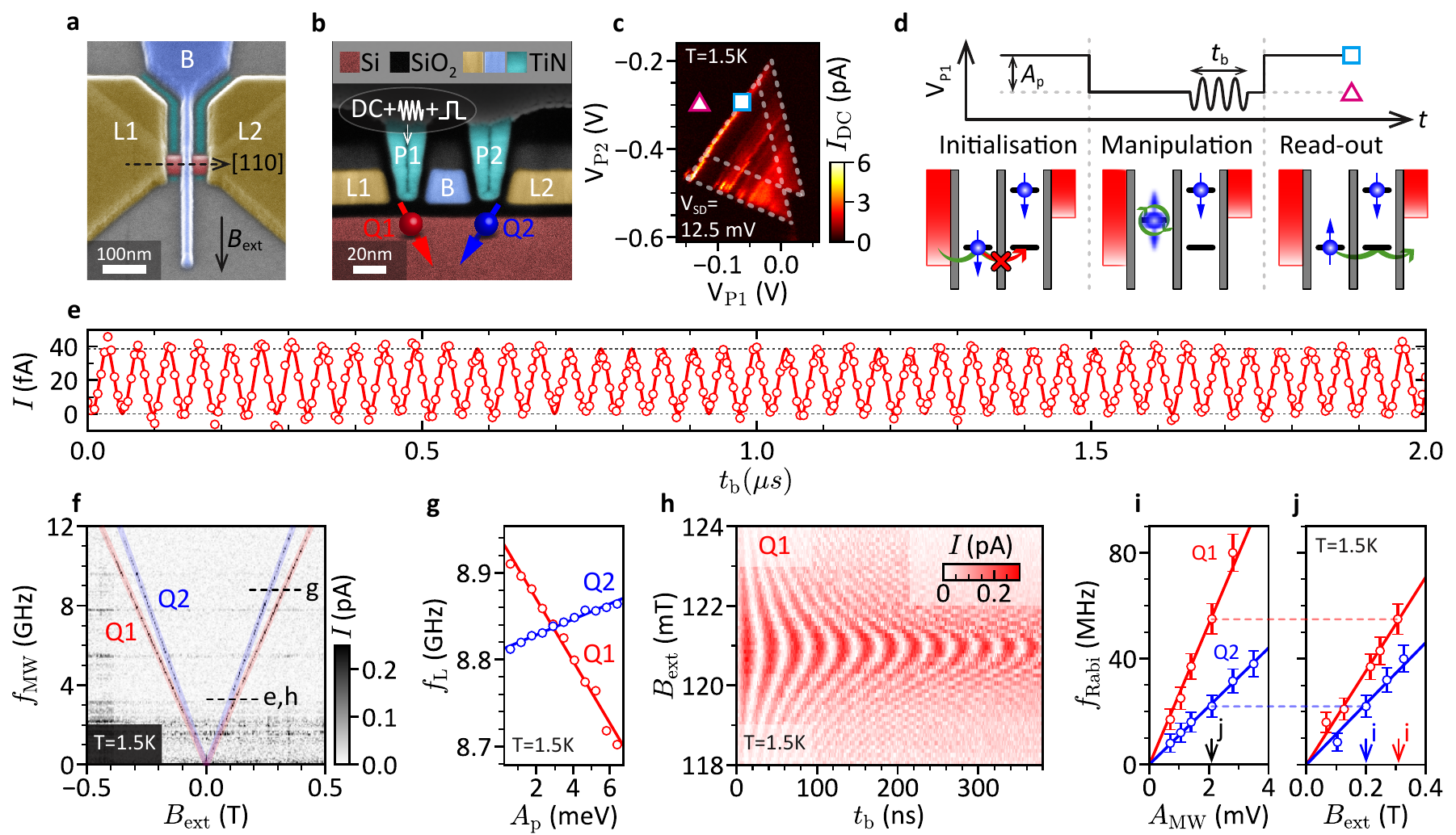}
\caption{\textbf{Spin-orbit qubits in a FinFET.} \textbf{a}, False-colour SEM image of an unfinished device showing the two lead gates L1, L2 (yellow) as well as the inter-dot barrier gate B (blue, ${\sim}\,\unit[35]{nm}$). An in-plane external magnetic field $B_{\mathrm{ext}}$ is applied perpendicular to the fin (red). \textbf{b}, Cross-sectional TEM image along the black dashed line shown in \textbf{a} after integration of the QDs' plunger gates P1, P2 (turquoise, ${\sim}\,\unit[15]{nm}$). In addition to a DC voltage, fast pulses and microwaves can be applied to P1. Current flow is observed from source to drain via the fin-shaped channel. \textbf{c}, Measurement of a spin-blocked pair of bias triangles. The blue square and pink triangle mark the qubit initialisation/readout and manipulation point, respectively. \textbf{d}, Schematic illustration of the spin manipulation cycle with corresponding pulse scheme. \textbf{e}, Rabi oscillation with $f_{\mathrm{Rabi}}=\unit[22]{MHz}$ measured on Q1 at $B_{\mathrm{ext}}=\unit[123]{\mathrm{mT}}$, $f_{\mathrm{MW}}=\unit[3.311]{GHz}$, $A_{\mathrm{MW}}=\unit[1.1]{mV}$  and $\mathrm{T}=\unit[1.5]{K}$. The data has been corrected by removing a small constant offset, and is fitted (solid curve) to $I(t_{\mathrm{b}})=A\sin(2\pi f_{\mathrm{Rabi}} t_{\mathrm{b}}+\theta)+B$ with $A$, $B$, $f_{\mathrm{Rabi}}$ and $\theta$ as fit parameters. \textbf{f}, Measurement of the current as a function of $f_{\mathrm{MW}}$ and $B_{\mathrm{ext}}$. Along the red (blue) line the spin resonance condition is met for Q1 (Q2). For each frequency the average current has been subtracted. \textbf{g}, Electrical tunability of the qubit frequency with the depth of the Coulomb pulse. Solid lines represent linear fits to the data. \textbf{h}, Detuned Rabi oscillations showing a typical chevron pattern, measured at $f_{\mathrm{MW}}=\unit[3.311]{GHz}$ and $A_{\mathrm{MW}}=\unit[1.4]{mV}$. Dependence of $f_{\mathrm{Rabi}}$ on $A_{\mathrm{MW}}$ \textbf{j} and $B_{\mathrm{ext}}$ \textbf{i}. Solid lines are linear fits to the data with zero offset.}
\label{fig1}
\end{figure}

A scanning electron microscope (SEM) tilted side-view and a transmission electron microscope (TEM) cross-sectional view of a co-fabricated device are shown in Figs.\ 1a,b. Since these FinFETs are fabricated using CMOS processes, they feature a highly uniform gate profile \cite{Zwerver2021} and ultra-small gate lengths \cite{Geyer2021}, resulting in an estimated effective dot size of \unit[{$\sim$}\,{7}]{nm} (Supplementary Section 6). By negatively biasing the gate electrodes, an accumulation-mode hole double quantum dot (DQD), hosting two individual spin-1/2 qubits, is formed \cite{Geyer2021}. Here, a pseudospin of $\pm1/2$ is assigned to the two lowest energy hole states, which for 1D-like hole systems can have large contributions of both heavy-hole and light-hole basis states \cite{Kloeffel2018,Bosco2021}. We measure the direct current $I_{\mathrm{DC}}$ through the DQD, which when combined with spin-to-charge conversion through Pauli spin blockade (PSB) \cite{Ono2002,Li2015} provides qubit readout functionality (for further details on the device and measurement setup see Methods). For the device investigated, PSB is observed for the $(1, 1)\rightarrow(0,2)/(2,0)$ charge state transitions and no additional transitions are observed when further depleting the QDs. Here $(m,n)$ denotes the effective hole occupancy of the left/right QD, while the true hole occupancy is $(m+m_0,n+n_0)$ with possible additional holes $m_0,n_0$.

In the PSB regime hole tunnelling is forbidden by spin conservation if the two spins, one per QD, occupy a spin-polarized triplet state ($\ket{(1,1)\mathrm{T}_+}$ or $\ket{(1,1)\mathrm{T}_-}$) and are thus aligned parallel. The unpolarized triplet $\ket{(1,1)\mathrm{T}_0}$ is not blockaded as it mixes with the singlet $\ket{(1,1)\mathrm{S}}$, which itself is coupled to the singlet $\ket{(0,2)\mathrm{S}}$ such that hole transport occurs \cite{Seedhouse2021}. Spin blockade can be lifted by flipping the direction of one hole spin using EDSR \cite{Nowack2007,NadjPerge2010,Maurand2016,Watzinger2018,Froning2021}, which is performed by applying square voltage pulses and microwave (MW) bursts to gate P1 (Fig.\ 1b). The measurements consist of three stages  (Figs.\ 1c,d): first, the two holes spins are initialised in a polarised spin state through PSB. Then, the system is pulsed into Coulomb blockade, where the MW signal is applied. Finally, in the readout stage a current is detected if the spins are antiparallel, such that one hole can tunnel to the neighbouring QD and exit to the nearby reservoir. This cycle is repeated many times for a measurable current, such that the duration of the manipulation stage is limited to a few microseconds (Supplementary Section 1).

For high-temperature operation of spin qubits \cite{Petit2020,Yang2020}, spin-to-charge conversion via PSB rather than energy-selective tunnelling \cite{Elzerman2004} is favourable, since the single-dot singlet-triplet splitting \cite{Geyer2021} is typically much larger than the Zeeman energy. Thus, the measurements can be performed at higher temperature and smaller external magnetic field, resulting in lower and technically less demanding qubit frequencies.

EDSR takes place under the condition that the MW frequency $f_{\mathrm{MW}}$ equals the Larmor frequency $f_L=|g^*|\, \mu_B\,|B_{\mathrm{ext}}|/h$, where $g^*$ denotes the effective hole Land\'{e} $g^*$-factor along the magnetic field $B_{\mathrm{ext}}$ direction, $\mu_B$ Bohr's magneton and $h$ Planck's constant. In Fig.\ 1f the resonance appears as a V-shape that maps out $f_L$ in the $f_{\mathrm{MW}}$-$B_{\mathrm{ext}}$ plane. The single-hole spin resonance conditions differ slightly for the two qubits (Q1, Q2), making them individually addressable. From the slope of the current lines, we extract absolute values for the $g^*$-factor of $1.94\pm0.05$ and $2.35\pm0.05$, respectively. These two different values indicate a sensitivity to the local electric fields, which also provides an additional control knob for the $g^*$-factor, and thus the qubit frequency \cite{Veldhorst2014,Yoneda2017,Crippa2018,Hendrickx2020,Froning2021}. This is confirmed by Fig.\ 1g, where the $f_L$-dependence on the square pulse amplitude $A_\mathrm{p}$ is shown. 

When the MW drive is on resonance, the DQD current reveals Rabi oscillations as a function of the burst duration $t_\mathrm{b}$. An example of a $\unit[22]{MHz}$ Rabi oscillation, whose decay time is too long to be observed within \unit[87]{$\pi$} rotations, corresponding to the longest applicable $t_\mathrm{b}$, is given in Fig.\ 1d. For a detuned $f_{\mathrm{MW}}$ the qubit rotates around a tilted axis on the Bloch sphere, resulting in faster rotations of reduced contrast as demonstrated by the chevron pattern seen in Fig.\ 1h. The Rabi frequency $f_{\mathrm{Rabi}}$ increases linearly not only with the MW amplitude $A_{\mathrm{MW}}$ (Fig.\ 1i), but also $B_{\mathrm{ext}}$ (Fig.\ 1j) as expected for SOI-mediated spin rotations \cite{Golovach2006,Nowack2007,Froning2021,Voisin2015,Crippa2018}. For these measurements, $A_{\mathrm{MW}}$ is calibrated using the photon-assisted-tunnelling response (Supplementary Section 3) \cite{Nowack2007}. The maximum $f_{\mathrm{Rabi}}$ observed is $\unit[147]{MHz}$ (Supplementary Section 5), which corresponds to a spin-flip time of just $\unit[{\sim}\,3.4]{ns}$. Under the assumption that EDSR occurs due to a periodic displacement of the wave function as a whole, the $g^*$-factor is not modulated \cite{Crippa2018} and $f_{\mathrm{Rabi}}$ depends on the spin-orbit length $l_{\mathrm{SO}}$ \cite{Nowack2007}. We can therefore state an estimate for $l_{\mathrm{SO}}$ in the range of $20$ to $\unit[60]{nm}$ (Supplementary Section 7), that is, similar values to the one reported before \cite{Geyer2021} and in very good agreement with theory predictions \cite{Kloeffel2018}.

A key parameter for the qubit controllability is the quality factor defined as $\mathrm{Q}=2f_{\mathrm{Rabi}}T_2^{\mathrm{Rabi}}$, where $T_2^{\mathrm{Rabi}}$ is the decay time of the Rabi oscillations. For the data presented in Fig.\ 1e no decay is observed within $\unit[{\sim}\,2]{\mu s}$, that is, $\mathrm{Q}\gg87$. In terms of quality factors, our hole spin qubits therefore outperform their hot electron counterparts \cite{Petit2020,Yang2020}.

\begin{figure}
\centering
\includegraphics{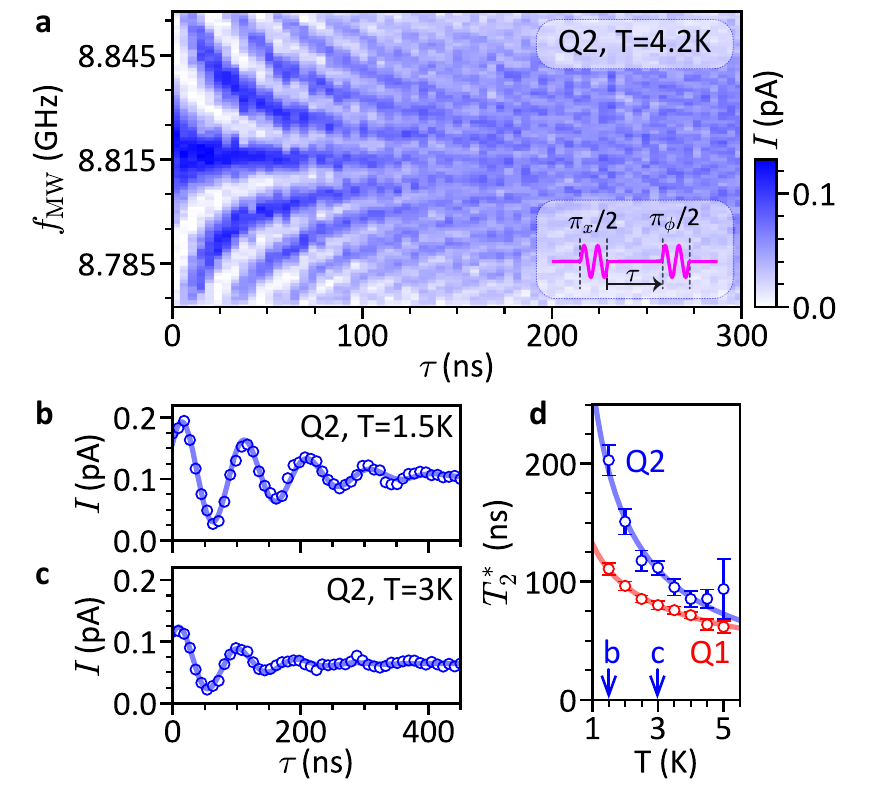}
\caption{\textbf{Hot qubit coherence. a,} Ramsey-fringe experiment performed at $\unit[4.2]{K}$. $B_{\mathrm{ext}}$ is fixed at $\unit[267]{\mathrm{mT}}$. The pulse sequence, which consists of two \unit[15]{ns}-long $\frac{\pi}{2}$-bursts separated by the waiting time $\tau$, is illustrated in the bottom right inset. $\phi$ denotes the phase of the second pulse with respect to the first one, here $\phi=0$. Decay of Ramsey fringes at $\unit[1.5]{K}$ \textbf{b} and $\unit[3]{K}$ \textbf{c}. The data were taken on resonance with a $\tau$-dependent phase $\phi(\tau)$, which adds an artificial oscillation \cite{Watson2018}. Solid curves show fits to $A+B\sin(\omega\tau+\theta)\exp[-(\tau/T_2^*)^{\beta+1}]$ with temperature dependent $\beta$. \textbf{d}, Temperature dependence of the spin dephasing time revealing a power-law decay $T_2^*\propto\mathrm{T}^{-\eta}$, where $\eta=0.46\pm0.02$ for Q1 and $\eta=0.81\pm0.06$ for Q2, respectively.}
\label{fig2}
\end{figure}

Next, we evaluate the spin coherence by performing a Ramsey experiment. Here, two $\frac{\pi}{2}$-pulses separated by a delay time $\tau$ during which the qubit can freely evolve and dephase are applied. When $f_{\mathrm{MW}}$ is detuned from the qubit resonance, the current through the device shows coherent oscillations known as Ramsey fringes. The data of Fig.\ 2a is measured at a temperature of $\mathrm{T}=\unit[4.2]{K}$, which corresponds to the boiling point of liquid $^4$He, and which can be achieved in a technically non-demanding way by immersing the sample in a liquid $^4$He bath or at the second stage of a dry pulse-tube refrigerator. The dephasing time $T_2^*$ is determined by fitting the envelope of the fringe decay to $\exp(-(\tau/T_2^*)^{\beta(\mathrm{T})+1})$, where $\beta$ depends on temperature as discussed later. Despite the fact that our qubit readout is protected against temperature by the large orbital energies, which exceed the thermal energy available at $\unit[4.2]{K}$ by an order of magnitude, a degradation of the signal contrast on increasing temperature is observed (Fig.\ 2b,c). The reasons for this are not yet fully understood, however we speculate that this is due to spin-flip cotunneling (Supplementary Section 8). The T-dependence of $T_2^*$ in the range of 1.5 to $\unit[5]{K}$ is presented for both qubits in Fig.\ 2d. While Q1 can be manipulated faster than Q2, it lags behind in coherence. The spin dephasing time drops with increasing temperature, described by a power-law decay $\propto\mathrm{T}^{-\eta}$ with $\eta=0.5$ (0.8) for Q1 (Q2), a rather weak temperature dependence similar to previous reports \cite{Yang2020,Petit2020}. The obtained values for $T_2^{*}$ are consistent with the EDSR spectral width (Supplementary Section 9), and a spin relaxation time ${T_1}\,\unit[{>}\,10]{\mu s}$ was found at \unit[4.2]{K} (Supplementary Section 12). In the following the focus is on the more coherent Q2.

Spin rotations around at least two different axes are required to reach any point on the Bloch sphere. In Fig.\ 3a we demonstrate two-axis qubit control at both $\unit[1.5]{K}$ and $\unit[4.2]{K}$ by employing a Hahn-type echo sequence. A modulation of the relative phase $\phi$ of the second $\frac{\pi}{2}$-pulse yields a set of Ramsey fringes that are phase-shifted by $\pi$ for a $\pi_x$ and $\pi_y$ echo pulse, which is applied to extend the coherence. The performance of the hole spin rotations is characterised using randomised benchmarking \cite{Knill2008,Muhonen2015} (see Fig.\ 3b and Methods). At $\unit[1.5]{K}$, a single-qubit gate fidelity of $F_{\mathrm{s}}=\unit[98.9\pm0.2]{\%}$ is obtained, which is at the fault-tolerance level \cite{Fowler2012,Veldhorst2014} and very similar to the values recently reported for hot electron spin qubits \cite{Yang2020,Petit2020}. The fidelity is reduced to $F_{\mathrm{s}}=\unit[98.6\pm1.6]{\%}$ ($\unit[97.9\pm1.1]{\%}$) at $\unit[3]{K}$ ($\unit[4.2]{K}$), revealing a similar scaling with temperature as $T_2^{*}$. We thus expect to be able to enhance the gate fidelities further by improving the qubit coherence, and by optimisation of the gate pulses \cite{Kelly2014}.

\begin{figure}
\centering
\includegraphics[width=\textwidth]{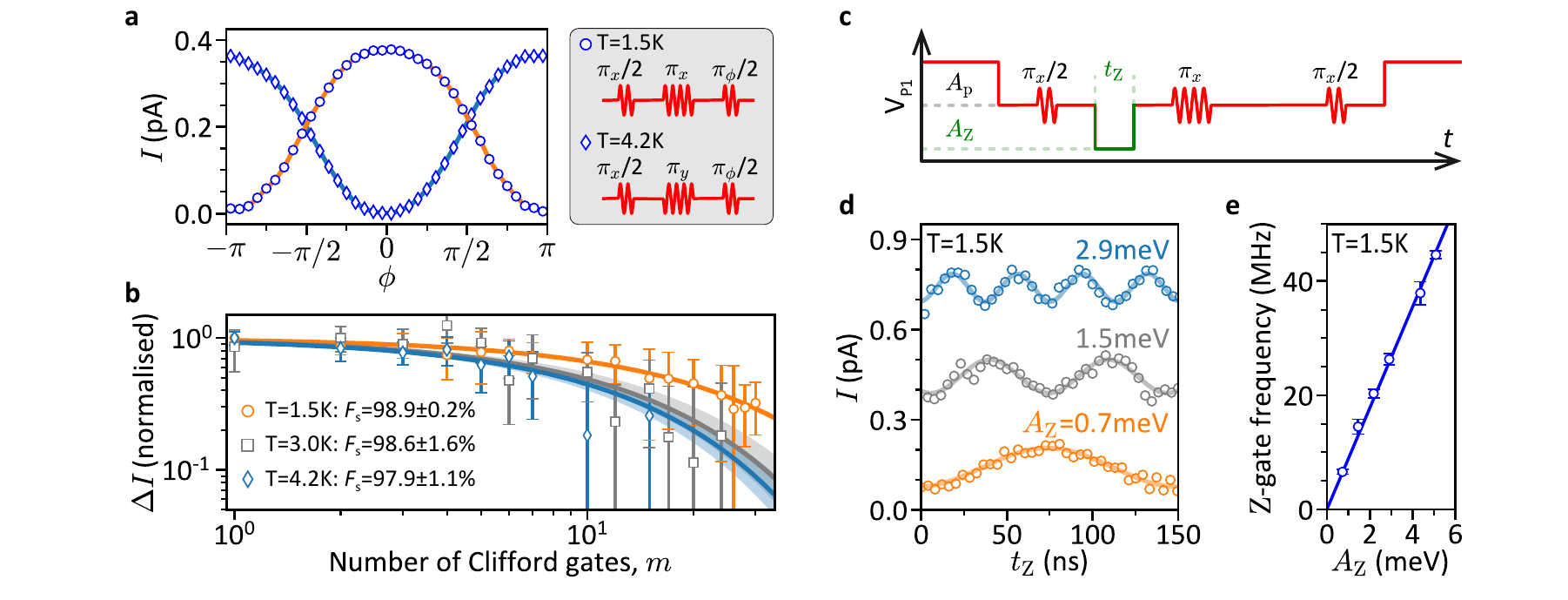}
\caption{\textbf{X, Y and Z qubit gates.} \textbf{a}, Demonstration of two-axis qubit control by applying a Hahn-type echo sequence, where the relative phase $\phi$ of the second $\frac{\pi}{2}$-pulse is varied. The measurements at $\unit[1.5]{K}$ (circles) and $\unit[4.2]{K}$ (diamonds) are phase-shifted by $\pi$ due to the two orthogonal echo pulses, as shown in the right panel.  \textbf{b}, Standard randomised benchmarking at $\unit[1.5]{K}$ (circles), $\unit[3]{K}$ (squares) and $\unit[4.2]{K}$ (diamonds) is performed by applying a varying number of Clifford gates $m$ and preparing either a $\ket{\uparrow}$ or $\ket{\downarrow}$ final state.  The normalised difference of currents is fitted to a single exponential decay to extract the single-qubit gate fidelities $F_\mathrm{s}$ (see Methods for further details). The shaded regions show the one-sigma error range of the fit parameters. The maximum $m$ decreases with increasing temperature due to a reduced readout contrast. \textbf{c}, Schematic representation of the pulse scheme used to demonstrate qubit rotations around the $z$-axis of the Bloch sphere. In a modified Hahn echo sequence a square pulse of amplitude $A_\mathrm{Z}$ and duration $t_{\mathrm{Z}}$ is applied to shift the qubit precession frequency (see Fig.\ 1\,g). The resulting phase-shift-induced oscillations are shown in \textbf{d} for different $A_\mathrm{Z}$. Solid curves represent fits to a sinusoidal function, where the oscillation frequency is given by the induced qubit frequency shift. Traces are offset by an increment of 0.3 for clarity. \textbf{e}, The speed of the $z$-rotations increases linearly with $A_\mathrm{Z}$. The solid line represents a linear fit to the data, yielding a frequency-shift of \unit[8.9]{MHz/meV}. The data presented in this figure was taken for Q2 at $f_{\mathrm{MW}}=\unit[8.812]{GHz}$.}
\label{fig3}
\end{figure}

Besides rotations around the $x$- and $y$-axis of the Bloch sphere, $z$-rotations can be realised by exploiting the electrical tunability of the qubit frequency (Fig.\ 1g). For this purpose a square pulse of amplitude $A_{\mathrm{Z}}$ and duration $t_{\mathrm{Z}}$ is added to a Hahn echo sequence (Fig.\ 3c) in order to rapidly detune the spin precession frequency, which leads to a phase pick up around the $z$-axis of the Bloch sphere \cite{Yoneda2017}. As a consequence, the DQD current oscillates as a function of $t_{\mathrm{Z}}$ (Fig.\ 3d) at a frequency that increases linearly with $A_{\mathrm{Z}}$ up to $\unit[{\sim}\,45]{MHz}$ (Fig.\ 3e).

\begin{figure}
\centering
\includegraphics{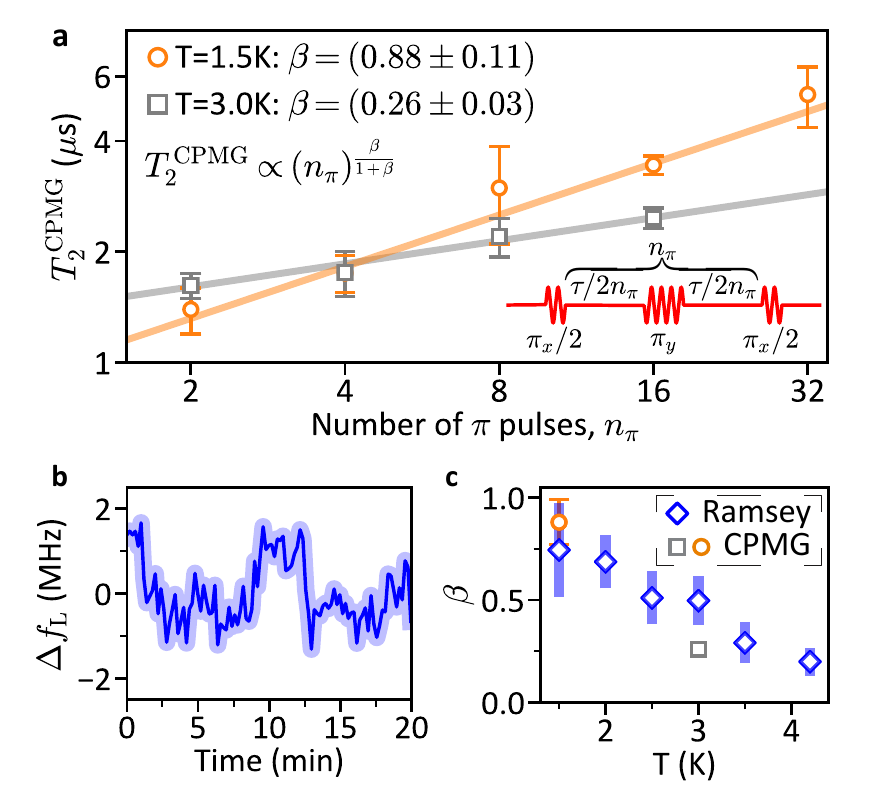}
\caption{\textbf{Dynamical decoupling and noise spectroscopy.} \textbf{a}, The spin coherence time can be enhanced by decoupling the qubit from low-frequency noise using a CPMG pulse sequence (see bottom-right schematic). A power-law dependence of the coherence time on the number of refocusing pulses $n_{\pi}$ is confirmed by fitting (solid lines) the data to $T_2^{\mathrm{CPMG}} = T_2^0(n_\pi)^{\frac{\beta}{1+\beta}}$, where $\beta$ represents the scaling exponent of a power-law noise spectrum, $S(f)\propto f^{-\beta}$. \textbf{b}, Time trace of the qubit frequency obtained from repeated Ramsey measurements. The shaded region indicates the frequency uncertainty due to readout noise. \textbf{c}, Temperature dependence of the noise exponent $\beta$ extracted from either CPMG or Ramsey measurements. The data presented in this figure was taken for Q2 at $f_{\mathrm{MW}}=\unit[8.812]{GHz}$.}
\label{fig4}
\end{figure}

Finally, in order to gain insight into the sources of decoherence we perform noise spectroscopy by employing Carr-Purcell-Meiboom-Gill (CPMG) pulse sequences \cite{Meiboom1958}, where a series of $n_{\pi}$ $\pi_y$-pulses is applied as a spectral filter for the environmental noise \cite{Bylander2011,Medford2012,Yoneda2017}. For a power-law noise spectrum $S(f) \propto f^{-\beta}$, the CPMG coherence time $T_2^{\mathrm{\,CPMG}}$ is expected to scale as $T_2^{\mathrm{\,CPMG}}\propto(n_{\pi})^{\frac{\beta}{1+\beta}}$ \cite{Medford2012}. This dependency is confirmed by Fig.\ 4a, and a $\beta$ of $0.88\pm0.11$ ($0.26\pm0.03$) is determined for $\unit[1.5]{K}$ ($\unit[3]{K}$), revealing a whitening of the noise on increasing the temperature and thus a reduced noise-decoupling efficiency. For $n_{\pi}=32$, the maximum $n_{\pi}$ achievable with our transport-based readout scheme, the hole spin coherence time is extended to $\unit[5.4]{\mu s}$ at $\unit[1.5]{K}$, which corresponds to an increase by a factor of 27 compared to the unprotected qubit. While our CPMG measurements are sensitive to the noise at frequencies of $f\unit[{\sim}\,{10^5}\,{-}\,10^7]{Hz}$, we independently probe $S(f)$ at $f\unit[{\sim}\,{10^{-3}}\,{-}\,10^{-1}]{Hz}$ by tracking the Larmor frequency fluctuations through repeated Ramsey experiments \cite{Yoneda2017} (Fig.\ 4b). The temperature dependence of $\beta$ demonstrates a noise whitening in both frequency ranges, and the good agreement of the $\beta$-values for the two frequency windows suggests a similar coloured noise spectrum over a wide range of frequencies. From the scaling of $\beta$ with T we cannot uniquely identify the underlying noise sources, such as charge or nuclear spin fluctuations \cite{Kuhlmann2013}. We note, however, that the longest $T_2^*$ measured is $\unit[{\sim}\,440]{ns}$ (Supplementary Section 10), which does not only exceed the dephasing times reported so far for hole spins in silicon at mK temperatures \cite{Hutin2018}, but is also close to the estimated limit of $\unit[{\sim}\,500]{ns}$ set by the hole spin hyperfine interaction (Supplementary Section 11). This sub-$\mu$s limit is a consequence of the hole spins interacting with a relatively small number of nuclear spins $N_s\sim310$, which increases the Overhauser field fluctuations that scale with $1/\sqrt{N_s}$ \cite{Assali2011}, and also represents a lower bound due the anisotropy of the hole hyperfine interaction \cite{Prechtel2016}.

We have reported hole spin qubits in silicon FinFETs that operate above $\unit[4]{K}$. The strong SOI allows for spin rotations as fast as $\unit[147]{MHz}$ and the weak hyperfine coupling ensures $T_2^*$ up to $\unit[440]{ns}$. In addition to two-axis control, we implement fast z-rotations by employing the electrical tunability of the $g^*$-factor. At $\unit[1.5]{K}$, we achieve fault-tolerant single-qubit gate fidelities. These results have been achieved using an industry-compatible FinFET device architecture, which is also well suited to implementing larger arrays of interacting qubits, such as a linear chain of exchange-coupled QD spins. Connectivity beyond nearest neighbours can be realised by coupling to a superconducting microwave resonator \cite{Borjans2019} or coherent spin shuttling \cite{Yoneda2021}. 

In the quest for a higher qubit quality factor, hyperfine-induced dephasing can be prevented by engineering a nearly nuclear-spin-free environment \cite{Veldhorst2014}. While a stronger SOI results in shorter gate times, it also increases the susceptibility to charge noise. For hole spins in silicon FinFETs, however, an unusually strong and at the same time electrically tunable SOI, allowing for on demand switching between qubit idling and manipulation modes, has been predicted \cite{Kloeffel2018,Froning2021, Bosco2021}. Furthermore, fast single-shot readout of hole spins is required for accurate qubit measurements. At few-kelvin temperatures this can be realised using a DQD charge sensor that exploits tunneling between two quantised states \cite{Huang2021}. This technique is more resilient against temperature than a single sensor QD, and high-fidelity single-shot readout up to \unit[8]{K} at a bandwidth greater than \unit[100]{kHz} was demonstrated. In addition, a read time resolution \unit[{$<$}\,{1}]{$\mu$s}, i.e.\ fast compared to our hole spin lifetime, was demonstrated using radio frequency reflectometry of a silicon DQD \cite{Noiri2020}. These readout techniques can be combined with the advance reported here, a hole spin qubit in a FinFET at temperatures of \unit[4]{K} and above. \\ 

\noindent\textbf{Methods}\\
\textbf{Device fabrication.} The fin structures are defined along [110] direction on a near-intrinsic, natural silicon substrate (${\rho>\unit[10]{k\Omega cm}}$, (100) surface) by means of electron-beam lithography (EBL) and dry etching \cite{Kuhlmann2018}. The gate oxide is formed by thermal oxidation of the silicon, yielding a $\unit[{\simeq}\,7]{nm}$-thick silicon dioxide (SiO$_2$) layer, which is covered by $\unit[{\simeq}\,20]{nm}$ of titanium nitride (TiN) grown by atomic layer deposition (ALD). The first layer of gates containing L1, L2 and B is patterned using EBL and dry etching. Subsequently, the gate stack ($\unit[{\simeq}\,4.5]{nm}$ SiO$_\mathrm{x}$, $\unit[{\simeq}\,20]{nm}$ TiN) of the second gate layer hosting P1 and P2 is grown by ALD. The plunger gates are implemented by means of a self-aligned process \cite{Geyer2021}, where the gaps between the gates of the first gate layer (highlighted in turquoise in Fig.\ 1a) act as a template for the plungers gates. The gate lengths of the device measured are $l_{\mathrm{B}}\,\unit[{\sim}\,35]{nm}$ and $l_{\mathrm{P}}\,\unit[{\sim}\,15]{nm}$. Source and drain contacts are p-type and made of platinum silicide (PtSi), which is formed by sputtering a $\unit[{\simeq}\,15]{nm}$-thick Pt layer on a beforehand cleaned silicon surface, followed by a silicidation anneal at $\unit[450]{^\circ C}$ for $\unit[10]{\mathrm{min}}$ in an argon ambient. Finally, the devices are encapsulated in a $\unit[{\simeq}\,100]{nm}$-thick SiO$_2$ layer that is grown by plasma-enhanced chemical vapour deposition and are accessed via tungsten interconnects.\\

\noindent\textbf{Experimental setup.} All measurements are performed using a variable temperature insert that can be operated at $\unit[1.5-50]{K}$. MW and DC signals can be applied simultaneously to gate P1 (see Fig.\ 1b) via a bias-tee on the sample board. DC voltages are supplied by a low-noise voltage source (BasPI SP927) and the source-drain current is measured with a current-to-voltage amplifier at gain $10^9$ (BasPI SP983c). A square voltage pulse used to drive the device between Coulomb blockade (qubit manipulation stage) and Pauli spin blockade (qubit initialisation and readout stage) is provided by an arbitrary waveform generator (Tektronix AWG5204), which also controls the I and Q inputs of a vector signal generator (Keysight E8267D) to generate phase-controlled square-shaped MW bursts. The latter ones and the square pulse are combined using a wideband power combiner (Mini-Circuits ZC2PD-5R263-S+). The qubit readout current is distinguished from the background by chopping the MW signal at a frequency of $\unit[89.2]{\mathrm{Hz}}$ and demodulating the current at this frequency with a lock-in amplifier (Signal Recovery 7265). For further details see Supplementary Section 1.\\

\noindent\textbf{Clifford benchmarking protocol.} Randomised benchmarking is performed by applying a randomised sequence of a varying number of Clifford gates $m$ before the spin state is rotated such that the final state ideally becomes either the $\ket{\uparrow}$ or $\ket{\downarrow}$ state. Each of the 24 gates in the Clifford group is constructed from the set $\lbrace \mathrm{I},\pm\mathrm{X}, \pm\mathrm{Y}, \pm\mathrm{X}/2, \pm\mathrm{Y}/2\rbrace$ \cite{Muhonen2015}. Assuming that the qubit initial state is $\ket{\downarrow}$, a current flow is only observed when spin blockade is lifted for a final $\ket{\uparrow}$ state. Thus, the difference in current between sequences designed to output either a $\ket{\uparrow}$ or $\ket{\downarrow}$ state, $\Delta\mathrm{I}=\mathrm{I}^{\ket{\uparrow}}-\mathrm{I}^{\ket{\downarrow}}$, is proportional to $p_{\uparrow}^{\ket{\uparrow}}-p_{\uparrow}^{\ket{\downarrow}}$. For each $m$ we average over 10 randomised sequences and the average Clifford-gate fidelity $F_\mathrm{c}$ is obtained from fitting the normalised current difference to $(2F_{\mathrm{c}}-1)^m$. Since a Clifford gate consists of on average 1.875 gates, the average single-qubit gate fidelity $F_{\mathrm{s}}$ is derived by $F_{\mathrm{s}}=1-(1-F_{\mathrm{c}})/1.875$.\\

\noindent\textbf{Data availability}\\
The data supporting the plots of this paper are available at the Zenodo repository at \url{https://doi.org/10.5281/zenodo.4579586.}\\

\noindent\textbf{Acknowledgements}\\
We thank M.~de Kruijf, C.~Kloeffel, D.~Loss, F.~Froning and F.~Braakman for fruitful discussions. Moreover, we acknowledge support by the cleanroom operation team, in particular by U.~Drechsler, A.~Olziersky and D.~Davila Pineda, at the IBM Binnig and Rohrer Nanotechnology Center, as well as technical support at the University of Basel by S.~Martin and M.~Steinacher. This work was partially supported by the Georg H. Endress Foundation, the NCCR SPIN, the Swiss Nanoscience Institute (SNI), the Swiss NSF (grant nr.\ 179024), and the EU H2020 European Microkelvin Platform EMP (grant nr.\ 824109). L.C.C. acknowledges support by a Swiss NSF mobility fellowship (P2BSP2\_200127).\\

\noindent\textbf{Author contributions}\\
A.V.K., L.C.C., S.G., A.F., R.J.W. and D.M.Z. conceived the project and experiments. A.V.K. and S.G. fabricated the device. L.C.C. and D.M.Z. prepared the cryogenic measurement setup. A.V.K. S.G., L.C.C. and D.M.Z. performed the experiments. A.V.K, L.C.C., and S.G. analysed the data and wrote the manuscript with input from all the authors.\\

\noindent\textbf{Competing interests}\\
The authors declare no competing interests.\\

\end{document}